\documentstyle[12pt]{article}
\begin{document}

\renewcommand{\baselinestretch}{1.5}
\newcommand\beq{\begin{equation}}
\newcommand\eeq{\end{equation}}

\centerline{\bf A comparative study of the phase diagrams of spin-$1 \over 2$ 
and spin-$1$} 
\centerline{\bf antiferromagnetic chains with dimerization and frustration}
\vskip 1 true cm

\centerline{Swapan Pati$^1$, R. Chitra$^2$, Diptiman Sen$^{3,5}$,}
\centerline{S. Ramasesha$^{1,5}$ and H. R. Krishnamurthy$^{4,5}$} 

\vskip .5 true cm

\centerline{\it $^1$ Solid State and Structural Chemistry Unit,} 
\centerline{\it Indian Institute of Science, Bangalore 560012, India}

\centerline{\it $^2$ Laboratoire de Physique des Solides, Universite 
Paris-Sud,}
\centerline{\it Batiment 510, 91405 Orsay, France}

\centerline{\it $^3$ Centre for Theoretical Studies, Indian Institute 
of Science,}
\centerline{\it Bangalore 560012, India}

\centerline{\it $^4$ Physics Department, Indian Institute of Science,}
\centerline{\it Bangalore 560012, India}

\centerline{\it $^5$ Jawaharlal Nehru Centre for Advanced Scientific 
Research,}
\centerline{\it Indian Institute of Science Campus, Bangalore 560012, India} 

\vskip 2 true cm

\leftline{\bf Abstract}
\vskip 1 true cm

We use the density matrix renormalization group method to study the ground 
state `phase' diagram and some low-energy properties of isotropic 
antiferromagnetic spin-$1 \over 2$ and spin-$1$ chains with a next-nearest 
neighbor exchange $J_2 ~$ and an alternation $\delta$ of the nearest neighbor 
exchanges. In the spin-$1 \over 2$ chain, the system is gapless for $\delta=0$ 
and $J_2 < J_{2c} =0.241$, and is gapped everywhere else in the $J_2 - \delta$ 
plane. At $J_{2c}$, for small $\delta$, the gap increases as $\delta^{\alpha}$, 
where $\alpha = 0.667 \pm 0.001$. $2J_2 + \delta = 1$ is a disorder line. To 
the left of this line, the structure factor $S(q)$ peaks at $q_{max} = \pi$ 
(Neel `phase'), while to the right, $q_{max}$ decreases from $\pi$ to $\pi/2$ 
(spiral `phase') as $J_2$ increases. There is also a `$\uparrow \uparrow 
\downarrow \downarrow$ phase' for large values of both $J_2$ and $\delta$. In 
the spin-$1$ case, we find a line running from a gapless point at $(J_2 ,
\delta) = (0,0.25 \pm 0.01)$ upto a `gapless' point at $(0.73 \pm 0.005,0)$ 
such that the open chain ground state is four-fold degenerate below the line 
and is unique above it. There is a disorder line in this case also and it has 
the same equation as in the  spin-$1 \over 2$ case, but the line ends at about 
$\delta =0.136$. Similar to the spin-$1 \over 2$ case, to the left of this 
line, the peak in the structure factor is at $\pi$ (Neel `phase'), while to 
the right of the line, it is at less than $\pi$ (spiral `phase'). For 
$\delta =1$, the system corresponds to a spin ladder and the system is gapped 
for all values of the interchain coupling for both spin-$1 \over 2$ and 
spin-$1$ ladders. 

\vskip 1 true cm


\newpage

\centerline{\bf I. INTRODUCTION}
\vskip .5 true cm

While the spin-$1 \over 2$ Heisenberg antiferromagnetic chain 
has been extensively studied using a variety of analytical and numerical 
techniques \cite{TON1}, the corresponding spin-$1$ chain has been studied in 
much less detail \cite{WHI1,TOT,TON2,SIN}. Interest in spin-$1$ chains grew 
after Haldane 
conjectured that integer spin chains with a nearest-neighbor (nn) 
exchange should have a gap while half-integer spin chains should be 
gapless. This observation was based on a non-linear sigma model (NLSM) 
field theory description of the low-energy excitations \cite{HAL}. The NLSM 
approach can be generalized to include other features such as 
dimerization (an alternation $\delta$ of the nn exchanges) and a 
next-nearest-neighbor (nnn) exchange $J_2 ~$ \cite{AFF1}, and it leads to 
interesting predictions. For the spin-$1 \over 2$ model, it predicts
that the system should be gapless for $J_2< J_{2c}$, for $\delta =0$
and should be gapped for all nonzero $\delta$. On the other hand, the theory
predicts that 
the spin-$1$ model should exhibit a gapless line in the $J_2 -\delta$ plane 
for nonzero $\delta$. If the nnn exchange 
is large enough, the spin chains go over from a Neel `phase' \cite{FN1} 
to a spiral `phase' and a different kind of NLSM field theory becomes 
applicable \cite{RAO1,ALL} which predicts a gap for {\it all} values of the 
spin. 

Real spin-$1 \over 2$ Heisenberg systems with both dimerization and 
frustration are
now known \cite{CAV}. However, the spin-$1$ analogs are yet to be synthesized.
In what follows, we demonstrate that the spin-$1$ system exhibits a richer 
`phase' diagram than the spin-$1 \over 2$ system. 
It is hoped that this will provide motivation for 
experimental realizations of such higher spin systems.

In this paper, we present a detailed comparative study of spin-$1$ and
spin-$1 \over 2$ chains with both dimerization and 
frustration in the $J_2 -\delta$ plane using the density matrix 
renormalization group (DMRG) method 
\cite{WHI2,CHI,PAT}. The major surprise which we discover is a `gapless' (to 
numerical
accuracy) point at $(J_2 =0.73, ~\delta =0)$, in the spin-$1$ case, which is 
contrary to the field theory expectation. 
We suggest that this point may be close to a critical point which is 
described by a $SU(3)$ symmetric conformal field theory \cite{SUT,AFF2}. 

\newpage

\centerline{\bf II. THE DMRG METHOD AND THE `PHASE' DIAGRAM}
\vskip .5 true cm

We have studied both open and periodic chains with an even number of 
sites governed by the Hamiltonian
\beq
H ~=~ \sum_i ~[ ~1 ~-~ (-1)^i ~\delta ~] ~ {\bf S}_i \cdot 
{\bf S}_{i+1} ~ + ~J_2 ~\sum_i ~{\bf S}_i \cdot {\bf S}_{i+2} ~, 
\label{ham1}
\eeq
with the limits of summation being interpreted as appropriate.  We restrict 
our attention to the region $J_2 ~\ge 0$ and $0 \le \delta \le 1$. We study 
various regions in the $J_2 - \delta$ plane using the DMRG method. The 
interactions are schematically shown in Fig. 1.

The DMRG technique involves systematically building up the chain to a 
desired number of sites starting from a very short chain by adding two sites
at a time. The initial chain of $2n$ sites (with $n$ a small enough integer) 
is diagonalized exactly. The reduced density matrix for the left $n$ sites 
is computed from the target state of the $2n$ chain Hamiltonian by 
integrating over the states of the right $n$ sites. This density matrix 
is diagonalized, and a matrix representation of the $n$-site Hamiltonian 
is obtained in a truncated basis with $m$ basis vectors which are the 
eigenvectors of the density matrix corresponding to its $m$ largest 
eigenvalues. The Hamiltonian matrix for the $2n+2$ chain is then obtained 
in the $(2s+1)^2 m^2 ~$ dimensional direct product subspace constructed using 
the truncated basis of the left and the right halves of the $2n$ chain and 
the full space of the two additional spins which are inserted in the middle. 
After obtaining the target state of the $2n+2$ chain in the truncated basis, 
the density matrix of half the chain, now with $n+1$ sites, is computed. 
The procedure is repeated till one reaches the desired chain length $N$. 
This algorithm is suited for studying an infinite system by extrapolation.
There is a different algorithm suited to studying a system with a specific 
finite size $N$ \cite{WHI2}; this iteratively improves the density 
matrices so that these matrices correspond to the actual density matrices
of the target state of the corresponding fragments in the desired finite 
size chain. This finite 
size algorithm could also be used for a detailed study of a specific finite
chain system. We have however chosen to extrapolate to the thermodynamic limit
by employing the algorithm described earlier for the infinite system.

The DMRG method allows us to study a few low-lying states in a sector
with a given value of the total spin component, $S_z$. The ground state 
is always the first (lowest energy) state in the $S_z = 0$ sector. The 
accuracy of the DMRG method depends crucially on the number of eigenstates
of the density matrix, $m$, which are retained. We have worked with $m=100$ to 
$120$ over the entire $J_2 - \delta$ plane after checking that the DMRG 
results obtained using these values of $m$ agree well with exact numerical 
diagonalizations of chains with upto $16$ sites for spin-$1$ \cite{TON2}
and $22$ sites for spin-$1 \over 2$ \cite{RAM}. The chain lengths we studied 
varied from $150$ sites for $J_2 > 0$ to $200$ sites for $J_2 =0$. We tracked 
our results as a function of $N$ and found that convergence is reached 
well before $150$ sites in all cases. We 
find that the numerical results are much 
better convergent for open chains than for periodic chains, a feature generic 
to the DMRG technique \cite{WHI2,SOR}. Hence the data shown in Figs. 2 
to 8, particularly for spin-$1$ chains, are mainly based on open chain results.

The `phase' diagrams which we obtain for spin-$1 \over 2$ and spin-$1$ chains
are shown in Figs. 2 and 3, respectively. In the spin-$1 \over 2$ case, the 
system is
gapless from $J_2 =0$ to $J_{2c} =0.241$ for $\delta =0$, and is gapped 
everywhere else in the $J_2- \delta$ plane. There is a disorder line, 
$2J_2 +\delta =1$, such 
that the peak in the structure factor $S(q)$ is at $q_{max} =\pi$ to the left
of the line, and decreases from $\pi$ to $\pi /2$ with increasing $J_2$ to the
right of the line (Fig. 4). Further, the correlation length $\xi$ goes 
through a minimum on this line. (We have borrowed the term `disorder line'
from the language of classical statistical mechanics \cite{SEL}).

In the spin-$1$ case (Fig. 3), the phase diagram is more complex. 
There is a solid
line marked $A$ which runs from $(0,0.25)$ to about $(0.22 \pm 0.02, 0.20 \pm 
0.02)$ shown by a cross. Within our numerical accuracy, the gap is zero on this 
line and the correlation length $\xi$ is as large as the system size $N$. The 
rest of the `phase' diagram is gapped. However the gapped portion can be 
divided into different regions characterized by other interesting features. On 
the dotted lines marked $B$, the gap is finite. Although $\xi$ goes through a 
maximum when we cross $B$ in going from region II to region I or from region 
III to region IV, its value is much smaller than $N$. There is a dashed line 
$C$ extending from $(0.65,0.05)$ to about $(0.73,0)$ on which the gap appears 
to be zero (to numerical accuracy), and $\xi$ is very large but not as large 
as $N$. In regions II and III, 
the ground state for an {\it open} chain has a four-fold 
degeneracy (consisting of $S=0$ and $S=1$), whereas it is nondegenerate in 
regions I and IV with $S=0$. The dashed line marked $D$ is defined by $2 J_2 + 
\delta = 1$, has an exactly dimerized ground state, and extends from $(0,1)$ to 
about $(0.432,0.136)$. The line $E$ separating regions II and III begins at 
about $(0.39,0)$ and extends upto the region V.  In regions I and II, the peak 
in the structure factor is at $\pi$ (Neel), while in regions III and IV, the 
structure factor peaks at less than $\pi$ (spiral). We 
will comment on all these features of the `phase' diagrams below.

\vskip .5 true cm
\centerline{\bf A. The frustrated spin chain (the line $\delta =0$)}
\vskip .5 true cm

For spin-$1 \over 2$, the system is gapless and has a unique ground state for 
weak frustration, i.e., $0 < J_2 < J_{2c} =0.241$. Beyond $J_{2c}$, the 
system is gapped and has two ground states \cite{CHI}; these are 
spontaneously dimerized \cite{AFF3}.

For spin-$1$, the system is gapped for all $J_2$ except for the `gapless'
point at $(0.73,0)$. For reasons explained in Sec. III, this `gapless' point  
is quite unexpected. So we examine that point in more detail. Fig. 5 shows 
a plot of the gap versus $J_2 ~$ for $\delta =0$. 
It is non-monotonic and is `gapless' at about $J_2 =0.73$. In
regions II and III, i.e., for $J_2 \le 0.735$, the open chain ground state 
is found to be four-fold degenerate. By comparing the energies of the 
low-lying states in sectors with $S_z =0, 1$ and $2$, we find
that the four ground states have $S=0$ and $1$. We therefore define the gap 
as the energy difference between the first state in the $S_z =0$ sector and 
the {\it second} state with $S_z =1$, since the gap to the first state with 
$S_z =1$ is zero. This is the correct definition of the gap since the 
finite ground state degeneracy arising from the end states (an artifact of the
open boundary conditions) does not contribute to thermodynamic 
properties. In region IV, i.e., for $J_2 > 0.735$, the ground state is found 
to be unique with $S=0$. So the gap is defined as the energy difference 
between the first states in the $S_z =0$ and $S_z =1$ sectors. In all cases,
we extrapolate the gap $\Delta$ to infinite system size by fitting it to $N$
using the form $\Delta = A + B/N^\alpha ~$, and finding the best possible
values of $A$, $B$ and $\alpha$ for each $J_2$ \cite{FN2}. 

Fig. 6 is a plot of the static structure factor $S(q)$ versus $q$ at four 
values of $J_2 ~$ in the neighborhood of $0.73$ obtained from open 
spin-$1$ chain studies with $150$ 
sites. For $J_2$ between $0.725$ and $0.735$, we see a pronounced peak at 
about $q_{max} = 112^o ~$. The peak decreases in height and becomes broader 
as one moves away from this interval.  We estimate the maximum value of $\xi$ 
to be about $60$ sites. It also decreases rapidly as we move away from that 
interval. Interestingly, Tonegawa {\it et al} \cite{TON2} did find a 
pronounced peak in $S(q)$ at $J_2 = 0.7$, although they did not investigate it 
further. 
 
It is natural to speculate that $(0.73,0)$ lies close to some critical point 
which exists in a bigger parameter space of spin-$1$ chains. We believe that 
the appropriate 
critical point may be the one discussed in Refs. \cite{SUT,AFF2}. Sutherland
exactly solves a spin-$1$ chain which has a nn biquadratic 
interaction of the form
\beq
H ~=~ \sum_i ~[ ~{\vec S}_i \cdot {\vec S}_{i+1} ~+~ \beta ~(~{\vec S}_i 
\cdot {\vec S}_{i+1} ~)^2 ~]~,
\label{ham2}
\eeq
with $\beta =1$, and finds that there are gapless modes at $q =0$ and 
$\pm ~120^o$ \cite{SUT}.
This implies a peak in the structure factor at $q =120^o ~$ which is not very 
far from the value we observe numerically. Affleck \cite{AFF2} further argues 
that the long-distance physics of this model is described by a conformal field
theory with $SU(3)$ symmetry \cite{FN3}.

\vskip .5 true cm
\centerline{\bf B. Ground state degeneracy}
\vskip .5 true cm

For spin-$1 \over 2$, the ground state is always unique except on the line
$\delta =0$ and $J_2 > 0.241$; for $J_2 > 0.241$, the ground state is 
two-fold degenerate.

For $\delta < 0.25$ and $J_2 =0$, the spin-$1$ chain is known to exhibit 
a `hidden' $Z_2 \times Z_2 ~$ symmetry breaking described by a non-local 
order parameter \cite{TOT,DEN}. 
This leads to a four-fold degeneracy of the ground state for the open 
chain. The degeneracy may be understood in terms of spin-$1 \over 2$ 
degrees of freedom living at the ends of the open chain whose mutual 
interaction decreases exponentially with chain length \cite{KEN}. We have 
oberved this ground state degeneracy at all points in regions II and
III in Fig. 2, where the gap between the singlet and triplet states vanishes 
exponentially with increasing chain length. In regions I and IV, 
the ground state is unique. The situation is reminiscent of the $Z_2 \times
Z_2 ~$ symmetry breaking mentioned above. However, we have not yet directly 
studied the non-local order parameter using the DMRG method. 

\vskip .5 true cm
\centerline{\bf C. Structure factor $S(q)$}
\vskip .5 true cm

We have examined the equal-time two-spin correlation function $C(r) =
\langle {\vec S}_o \cdot {\vec S}_r \rangle$, as well 
its Fourier transform $S(q)$. Since there is no 
long-range order anywhere in the $J_2 - \delta$ plane (except for algebraic
order on the lines $A$ in Figs. 2 and 3), $S(q)$ generally has a broad peak 
at some $q_{max} ~$. 

For spin-$1 \over 2$, $q_{max}$ is pinned at $\pi$ in region I (Neel), 
decreases 
from $\pi$ (near the straight line $B$) to $\pi /2$ (near the numerically found
curve $C$) in region II (spiral), and is pinned at $\pi /2$ in region III
($\uparrow \uparrow \downarrow \downarrow$). These
features are found by studying the behavior of $S(q)$ on all the points
marked in Fig. 2. We assign a point to the $\uparrow \uparrow \downarrow
\downarrow$ `phase' if the sign of
$C(r)$ alternates as $++--$ for $40$ consecutive sites in a $100$ site chain.

For spin-$1$, in regions I and II in Fig. 2, $q_{max} ~$ is pinned at 
$\pi$, while in regions III and IV, $q_{max} ~< \pi$. Above the curve $ABC$, 
the cross-over from the Neel to the spiral `phase' presumably occurs 
across the straight line $D$ given by $2 J_2 + \delta =1$ (see below). 
Below $ABC$, the cross-over has been determined purely numerically and seems 
to occur across the line indicated as $E$ in Fig. 2. The region of 
intersection between the cross-overs from Neel to spiral and from 
four-fold degeneracy to a unique ground state is a small `hole' (region
V) in the `phase' diagram centred about the point $(0.435,0.12)$. Points in 
this `hole' turned out to be extremely difficult to study using the DMRG method
because of convergence difficulties with increasing chain length. We have not
shown the $\uparrow \uparrow \downarrow \downarrow$ `phase' in the diagram 
for spin-$1$. However, we do find that the boundary of this phase for 
spin-$1$ is closer to the large-$S$ boundary (given below as 
$4J_2 =(1-\delta^2)/ \delta $) than for spin-$1 \over 2$.

\vskip .5 true cm
\centerline{\bf D. Disorder lines}
\vskip .5 true cm

For spin-$1 \over 2$, the straight line $B$ ($2 J_2 + \delta = 1$) indicated
in Fig. 2 can be shown to have a dimerized state as the exact ground
state. It is easy to show that a dimerized state of the form 
\beq
\psi ~=~ [1,2] ~[3,4]~.... ~[N-1 ,N]~,
\label{psi}
\eeq
where $[i,j]$ denotes the normalized singlet combination of the spins on 
sites $i$ and $j$, is an eigenstate of the Hamiltonian on that line. To prove 
that (\ref{psi}) is the ground state, we decompose the Hamiltonian as 
\beq
H ~=~ \sum_i ~H_i ~, 
\label{ham3}
\eeq
where each of the $H_i ~$ only acts on a cluster of $3$ neighboring sites. 
Next, we numerically show that (\ref{psi}) is a ground state of each of the 
$H_i ~$, and is therefore a ground state of $H$ by the Rayleigh-Ritz 
variational principle. 

For spin-$1$, the above proof that $\psi$ in Eq. (\ref{psi}) is the ground 
state holds only between $\delta =1$ and $\delta = 1/3$ \cite{SHA}, where 
each of the $H_i ~$ is a $3$-cluster Hamiltonian. For $\delta <1/3$ along
the disorder line, $\psi$ in (\ref{psi}) 
is no longer the ground state of any of the $3$-cluster Hamiltonians $H_i ~$. 
But we can construct $4$-cluster $H_i ~$ satisfying (\ref{ham3}) such that 
$\psi$ in (\ref{psi}) can be 
numerically shown to be a ground state of each of the $H_i ~$. This allows us 
to prove that $\psi$ in (\ref{psi}) is the ground state of 
$H$ upto a point which is 
further down the line $D$. By repeating this procedure with bigger and 
bigger cluster sizes $n$, we can show that $\psi$ in (\ref{psi}) is the 
ground state down 
to about $\delta =0.136$. At that value of $\delta$, the cluster size $n$ 
is as large as the largest system sizes that we have studied by the DMRG
method. Hence the argument that (\ref{psi}) is the ground state could not be 
continued further. The difficulty is augmented by the fact that below 
$\delta =0.136$, we have the `hole' (region V) where 
computations are not convergent. Since the segment of the straight line from 
the point $(0,1)$ upto the `hole' has an exactly known ground state with 
an extremely short correlation length (essentially, one site), and since 
there is a cross-over from a Neel to a spiral `phase' across the line, we
choose to
call it a disorder line just as in the spin-$1 \over 2$ case \cite{CHI}. 

Our DMRG studies show that the disorder line $D$ does not extend below 
the `hole' region; instead a new line $E$ emerges as the disorder line.
It is worthwhile noting that the line $E$ is found only numerically unlike
line $D$ which is obtained analytically.

\vskip .5 true cm
\centerline{\bf E. Coupled spin chains $(\delta =1)$}
\vskip .5 true cm

For $\delta =1$, we get two coupled spin chains as can be seen in Fig. 1;
the interchain 
coupling is $2$ and the intrachain coupling is $J_2$. We have scaled the
intrachain coupling to $1$, and have varied the interchain coupling $J$
in these scaled units. We have studied the
dependence of the gap $\Delta$ and the two-spin correlation function
$C(r)$ on the interchain coupling $J$. We have plotted $\Delta$ versus
$J$ for both spin-$1 \over 2$ and spin-$1$ in Fig. 7.

For spin-$1 \over 2$, we find that the system is gapped for any non-zero 
value of 
the interchain coupling $J$, although the gap vanishes as $J \rightarrow 0$.
We find that the gap increases and correspondingly the correlation length 
decreasess with increasing $J$.
 
For spin-$1$, we find the somewhat surprising result that both the gap and the 
correlation length $\xi$ are fairly large for moderate values of $J$. Note
that the variation of the gap with $J$ for spin-$1$ (shown as circles) 
is much less
than that for spin-$1 \over 2$ (crosses). Fig. 8 shows the correlation function
$C(r)$ as a function of $r$ for $J=0.286$ for spin-$1$. This data is for an 
open ladder with $150$ sites, and it is consistent with a value of $\xi$ 
which is much longer than that found in the spin-$1 \over 2$ case.

\vskip .5 true cm
\centerline{\bf III. NLSM FIELD THEORIES OF ANTIFERROMAGNETIC}
\centerline{\bf SPIN CHAINS}
\vskip .5 true cm

\centerline{\bf A. The $J_2 - \delta$ model}
\vskip .5 true cm

Briefly, the field theoretic analysis of spin chains with the inclusion
of $J_2 ~$ and $\delta$ proceeds as follows. In the 
$S \rightarrow \infty$ limit, a classical treatment (explained briefly in the
next subsection) shows that the ground 
state of the model is in the Neel phase for 
$4 J_2 < 1 - \delta^2$, in  a spiral phase for $1 - \delta^2 < 4 
J_2 < (1 - \delta^2) /\delta$, and in a `$\uparrow \uparrow \downarrow 
\downarrow$' phase for $(1 - \delta^2 ) /\delta < 4 J_2$ \cite{RAO2} (Fig. 9). 
These three phases differ as follows. In the 
classical ground state, all the spins can be shown to lie in a plane. Let
us define the angle between spins
${\vec S}_i$ and ${\vec S}_{i+1}$ to be $\theta_1$ if $i$ is odd and 
$\theta_2$ if $i$ is even. In the Neel phase, $\theta_1 = \theta_2 = \pi$.
In the spiral phase, $\theta_1 = \theta_2 = \cos^{-1} (-1/4J_2)$ if $\delta
=0$. In the $\uparrow \uparrow \downarrow \downarrow$ phase, $\theta_1 = 
\pi$ and $\theta_2 = 0$. 

To next order in $1/S$, one derives a 
semiclassical field theory to describe the long-wavelength low-energy 
excitations. The field theory in the Neel phase is given by a $O(3)$ 
NLSM with a 
topological term \cite{HAL,AFF1}. The field variable is a unit vector 
$\vec \phi$ with the Lagrangian density
\beq
{\cal L} ~=~ {1 \over {2 c g^2}} ~{\dot {\vec \phi}}^2 ~-~ {c \over {2 
g^2}} ~{\vec \phi}^{\prime 2} ~+~ {\theta \over {4 \pi}} ~{\vec \phi} 
\cdot {\vec \phi}^{\prime} \times {\dot {\vec \phi}} ~,
\label{lag1}
\eeq
where $c = 2 S (1 - 4J_2 - \delta^2 ~)^{1/2} ~$ is the spin wave 
velocity, $g^2 = 2 /[S (1 - 4 J_2 - \delta^2 ~)^{1/2} ]~$ is the coupling 
constant, and $\theta = 2 \pi S (1 - \delta)$ is the coefficient of the
topological term. Note that $\theta$ is independent of $J_2 ~$ in the 
NLSM. (Time and space derivatives are denoted by a dot and a prime 
respectively). For $\theta = \pi$ mod $2 \pi$ and $g^2 ~$ less than a 
critical value, the system is gapless and is described by a conformal field 
theory with an $SU(2)$ symmetry \cite{AFF1,AFF2}. For any other value 
of $\theta$, the system is 
gapped. For $J_2 = \delta =0$, one therefore expects that integer spin 
chains should have a gap while half-integer spin chains should 
be gapless. This is known to be true even for small values of $S$ like 
$1/2$ (analytically) and $1$ (numerically) although the field theory is 
only derived for large $S$. In the presence of dimerization, one expects 
a gapless system at certain special values of $\delta$. For $S=1$, the 
special value is predicted to be $\delta_c =0.5$. We see that the {\it 
existence} of a gapless point is correctly predicted by the NLSM. However, 
according to the DMRG results, $\delta_c~$ is at $0.25$ for $J_2 =0$ \cite{TOT} 
and decreases with $J_2 ~$ as shown in Fig. 3. These 
deviations from field theory are probably due to higher order corrections 
in $1/S$ which have not been studied analytically so far.
 
In the spiral phase, it is necessary to use a different NLSM which is known 
for $\delta =0$ \cite{RAO1,ALL}. The field variable is now an $SO(3)$ matrix 
$\underline R$ and the Lagrangian density is
\beq
{\cal L} ~=~ {1 \over {2 c g^2}} ~{\rm tr} ~\Bigl( ~{\dot 
{\underline R}^T} {\dot {\underline R}} ~P_0 ~\Bigr) ~-~ 
{c \over {2 g^2}} ~{\rm tr} ~\Bigl( ~
{\underline R}^{\prime T} {\underline R}^{\prime} ~P_1 ~\Bigr) ,
\label{lag2}
\eeq
where $c = S (1 + y) {\sqrt {1 - y^2}} / y$, $g^2 = 2 
{\sqrt {(1 + y) /(1 - y)}} /S$ with $1/y = 4 J_2 ~$, and $P_0 ~$ and 
$P_1 ~$ are diagonal matrices with diagonal elements $(1,1,2 y (1 -
y) / (2 y^2 - 2 y + 1) )$ and $(1,1,0)$ respectively. Note that
there is no topological term; indeed, none is possible since $\Pi_2 
(SO(3)) =0$ unlike $\Pi_2 (S^2) = Z$ for the NLSM in the Neel phase. 
Hence there is no apparent difference between integer and half-integer 
spin chains in the spiral phase. A one-loop renormalization group \cite{RAO1} 
and large $N$ analysis \cite{ALL} indicate that the system should have a gap 
for all values of $J_2 ~$ 
and $S$, and that there is no reason for a particularly small gap at any 
special value of $J_2 ~$. A similar conclusion is obtained from a bosonic 
mean-field theory analysis of the frustrated spin chain \cite{RAO3}. The 
`gapless' point at $J_2 =0.73$ for spin-$1$ is therefore surprising. 

In the $\uparrow \uparrow \downarrow \downarrow$ phase, the NLSM is known 
for $\delta =1$, i.e., for the spin ladder. The Lagrangian is the same as 
in (\ref{lag1}), with $c= 4S [J_2 (J_2 +
1)]^{1/2}$ and $g^2 = (1 + 1/J_2)^{1/2} / S$. There is {\it no} topological 
term for any value of $S$, and the model is therefore gapped.

Note that the `phase' boundary between Neel and spiral for spin-$1$ is 
closer to the the classical ($S \rightarrow \infty$) boundary $4J_2 = 1 - 
\delta^2$ than for spin-$1 \over 2$. For instance, the crossover from Neel to
spiral occurs, for $\delta =0$, at $J_2 = 0.5$ for spin-$1 \over 2$, at $0.39$ 
for spin-$1$, and at $0.25$ classically.

\vskip .5 true cm
\centerline{\bf B. The frustrated and biquadratic spin-$1$ models}
\vskip .5 true cm

For spin-$1$, there is a striking similarity between the ground state 
properties of our
model (\ref{ham1}) as a function of $J_2$ (with $\delta =0$) and the 
biquadratic model (\ref{ham2}) as a function of (positive) $\beta$ \cite{JOL}. 
For $J_2 < 0.39$ and $\beta < 1/3$, both models are in the Neel phase and
are gapped. For $J_2 > 0.39$ and $\beta > 1/3$, the two models are in the
spiral phase and are generally gapped; however, model (\ref{ham1}) is
`gapless' for $J_2 = 0.73$ while model ({\ref{ham2}) is gapless for $\beta =
1$. We can qualitatively understand the cross-over from the Neel to the
spiral phase (but {\it not} the gaplessness at a particular value of $J_2$
or $\beta$) through the following classical argument. Let us set the 
magnitudes of the spins equal to $1$ and define the angle between spins
${\vec S}_i$ and ${\vec S}_{i+n}$ to be $n \theta$. The angle $\theta$ can 
be obtained by minimizing $\cos \theta + J_2 \cos 2 \theta$ in (\ref{ham1}),
and $\cos \theta + \beta \cos^2 \theta$ in (\ref{ham2}). This gives us
a Neel phase ($\theta = \pi$) if $J_2 \le 1/4$ and $\beta \le 1/2$ in the
two models, and a spiral phase for larger values of $J_2$ and $\beta$
with $\theta = \cos^{-1} (-1/4 J_2)$ and $\theta = \cos^{-1} (-1/2 \beta)$
respectively. The actual crossover points from Neel to spiral is different for
spin-$1$ than these classical values.

\vskip .5 true cm
\centerline{\bf IV. SUMMARY}
\vskip .5 true cm

To conclude, we have studied a two-parameter `phase' diagram for the
ground state of isotropic antiferromagnetic spin-$1 \over 2$ and spin-$1$ 
chains. The spin-$1$ diagram is
considerably more complex than the corresponding spin-$1 \over 2$ chain 
with surprising features like a `gapless' point inside the 
spiral `phase'; this point could be close to a critical 
point discussed earlier in the literature \cite{SUT,AFF2}. 
It would be interesting to 
establish this more definitively. Our results show that 
frustrated spin chains with small values of $S$ exhibit features not 
anticipated from large $S$ field theories.

\vskip .5 true cm
\leftline{\bf Acknowledgments}
\vskip .5 true cm

We thank Sumathi Rao and B. Sriram Shastry for stimulating discussions, and 
Biswadeb Datta for assistance with the computer systems.

\newpage

\leftline{\bf Figure Captions}
\vskip .5 true cm

\begin{enumerate}

\item Schematic picture of the spin chain given by Eq. (\ref{ham1}).

\item `Phase' diagram for the spin-$1 \over 2$ chain in the $J_2 - \delta $
plane. The line $A$ from $(0,0)$ to $(0.241,0)$ is gapless; the rest of the
diagram is gapped. The straight line $B$ satisfying $2J_2 + \delta =1$ extends
all the way from $(0,1)$ to $(0.5,0)$. Across $B$, the position of the peak
in the structure factor decreases from $\pi$ (Neel) in region I to less than
$\pi$ (spiral) in region II. Across $C$, the peak
in the structure factor decreases from greater than $\pi /2$ (spiral) in 
region II to $\pi /2$ in region III ($\uparrow \uparrow \downarrow 
\downarrow$ `phase'). The two-spin correlation function and structure factor
were studied at all the points shown in the figure.

\item `Phase' diagram for the spin-$1$ chain. 
The solid line $A$ extending from $(0,0.25)$ upto the cross is gapless; the
rest of the diagram is gapped. On the dotted lines $B$, the gap is finite. The 
dashed line $C$ close to $(0.73,0)$ is `gapless'. The ground state for an 
open chain has a four-fold degeneracy in regions II and III, while it is 
unique in regions I and IV. The straight line $D$ satisfying $2 J_2 + \delta 
= 1$ extends from $(0,1)$ to about $(0.432, 0.136)$. Regions II and III are 
separated by line $E$ which goes down to about $(0.39,0)$. Across $D$ and $E$, 
the peak in the structure factor decreases from $\pi$ (Neel) 
in regions I and II to less than $\pi$ (spiral) in regions III and IV. 
The positions of all the points have an uncertainty of $\pm 0.01$ unless 
stated otherwise.

\item Plot of $q_{max}$ (in degrees) versus $J_2$ at $\delta = 0$ for
spin-$1 \over 2$.

\item Dependence of the gap on $J_2$ at $\delta =0$ for spin-$1$.

\item Structure factor $S(q)$ versus $q$ for $J_2 =0.71$, $0.72$, $0.725$ and 
$0.735$ at $\delta =0$ for spin-$1$.

\item Gap $\Delta$ versus $J$ for coupled spin chains ($\delta 
=1$). Spin-$1 \over 2$ and spin-$1$ data are indicated by crosses and circles
respectively.

\item Two-spin correlation function $C(r)$ versus $r$ for coupled spin-$1$
chains with $J = 0.286$.

\item Classical phase diagram of the spin chain in the $J_2 -\delta$ plane.

\end{enumerate}


\begin{thebibliography}{99}
\bibitem{FN1} We use the word `phase' only for convenience to distinguish 
between regions with different modulations of the two-spin correlation 
function. Our model actually has no phase transition from Neel to spiral even 
at zero temperature.
\bibitem{FN2} For $J_2 \le 0.735$, we also studied the
separation between the second and the third states with $S_z =1$. This
separation is appreciably larger than the extent of non-monotonicity of the
gap in the
region around $J_2 =0.5$. This confirms that the non-monotonic behaviour in
that region is not an artifact arising from a possible mixing up of the 
first and second excited states in our diagonalization algorithm.
\bibitem{FN3} At the $SU(3)$ symmetric critical point, the two-spin 
correlation should asymptotically decay as the $4/3$ power of the 
distance \cite{AFF2}. We tried to verify this but the finite correlation 
length prevented us from obtaining an accurate estimate of the power.
\end{thebibliography}
\end{document}